\pdfoutput=1

\documentclass[11pt]{article}

\usepackage[preprint]{coling}

\usepackage{times}
\usepackage{latexsym}
\usepackage{amsmath}
\usepackage{rotating}

\usepackage[T1]{fontenc}

\usepackage[utf8]{inputenc}

\usepackage{microtype}

\usepackage{inconsolata}

\usepackage{graphicx}
\usepackage{listings}
\newcommand{\summary}{\textit{summary}}
\newcommand{\reasoning}{\textit{reasoning}}
\newcommand{\factsingle}{\textit{fact\_single}}

\newcommand{\unanswerable}{\textit{unanswerable}}

\newcommand{\llama}[1]{\texttt{Llama-#1-70b}}

\newcommand{\llamalabels}{\texttt{LLama3}}

\newcommand{\squad}{\texttt{SQuAD2}}
\newcommand{\newsqa}{\texttt{NewsQA}}
\newcommand{\pubmedqa}{\texttt{PubMedQA}}
\newcommand{\hotpotqa}{\texttt{HotpotQA}}
\newcommand{\msmarco}{\texttt{MS MARCO}}
\newcommand{\naturalq}{\texttt{NaturalQ}}

\usepackage{booktabs, array}
\usepackage{cleveref}
\usepackage{multirow}
\title{Know Your RAG:\\Dataset Taxonomy and Generation Strategies for Evaluating RAG Systems}

\author{
 \textbf{Rafael Teixeira de Lima\textsuperscript{1*}},
 \textbf{Shubham Gupta\textsuperscript{1}},
\\
 \textbf{Cesar Berrospi\textsuperscript{2}},
 \textbf{Lokesh Mishra\textsuperscript{2}},
 \textbf{Michele Dolfi\textsuperscript{2}},
 \textbf{Peter Staar\textsuperscript{2}},
 \textbf{Panagiotis Vagenas\textsuperscript{2}}
\\
\\
 \textsuperscript{1}IBM Research Paris-Saclay, 2 Rue d’Arsonval, Orsay, France; \\
 \textsuperscript{2}IBM Research Zurich, Säumerstrasse 4, Rüschlikon, Switzerland;
\\
 \small{
   \textbf{\textsuperscript{*}Correspondence:} \href{mailto:rtdl@ibm.com}{rtdl@ibm.com}
 }
}

\begin{document}
\maketitle


\begin{abstract}
Retrieval Augmented Generation (RAG) systems are a widespread  application of Large Language Models (LLMs) in the industry. 
While many tools exist empowering developers to build their own systems, measuring their performance locally, with datasets reflective of the system's use cases, is a technological challenge. 
Solutions to this problem range from non-specific and cheap (most public datasets) to specific and costly (generating data from local documents). 
In this paper, we show that using public question and answer (Q\&A) datasets to assess retrieval performance can lead to non-optimal systems design, and that common tools for RAG dataset generation can lead to unbalanced data. We propose solutions to these issues based on the characterization of RAG datasets through labels and through label-targeted data generation. Finally, we show that fine-tuned small LLMs can  efficiently generate Q\&A datasets. We believe that these observations are invaluable to the know-your-data step of RAG systems development.
\end{abstract}

\section{Introduction}
\label{section:introduction}

A Retrieval Augmented Generation (RAG) system pairs a Large Language Model (LLM) with an external knowledge source \citep{LewisEtAl:2020:RAGForKnowledgeIntensiveNLPTasks,GuuEtAl:2020:RALMPreTraining}. Given a user's query, a \emph{retriever} adds relevant information from the knowledge source (context) to the LLM's context window,  \emph{augmenting} the LLM's internal knowledge, and helping it \emph{generate} a grounded answer with fewer hallucinations \citep{PetroniEtAl:2020:HowContextAffectsLMsFactualPredictions}. This setup allows LLMs to use information like current news and private enterprise data that was not part of their training data \citep{IBM:2023:IBM-RAG}, which has prompted rapid adoption across the community \citep{NakanoEtAl:2021:WebGPT,ShusterEtAl:2022:BlenderBot3, SemnaniEtAl:2023:Wikichat,Nvidia:2023:ChatRTX}.

This adoption has been accompanied by a growing interest in strategies for evaluating RAG systems. Recent works focus on evaluating the entire system on a downstream task like question-answering \citep{ChenEtAl:2017:ReadingWikipediaToAnswerODQ}. Others separately measure the retriever's ability to fetch correct information \citep{KarpukhinEtAl:2020:DensePassageRetrievalForODQA,SalemiZamani:2024:EvaluatingRetrievalQualityInRAG} and the generator's ability to incorporate it in the output \citep{LiuEtAl:2023:RECALL,ChenEtAl:2024:BenchmarkingLLMsInRAG}. 
Tools like Ragas \citep{EsEtAl:2024:Ragas}, ARES \citep{SaadFalconEtAl:2024:ARES}, and LlamaIndex ~\cite{Liu_LlamaIndex_2022} have been developed for automated LLM-assisted evaluation of RAG systems. While these approaches focus on evaluation \emph{methods}, we take a step back in this paper and instead focus on the \emph{data} used for evaluation (i.e., a set of (context, query, answer) triplets). 


Our \textbf{first} contribution is a taxonomy for question-context pairs. We propose labels that identify different ways a user might interface with a RAG system on a given dataset. We show that popular public Q\&A datasets can be heavily unbalanced with respect to these labels, and that the performance of popular retrieval strategies can differ significantly across these classes. This can lead to performance measurements that do not reflect how users would interact with a given system, depending on what types of labels are expected in practice. 

Our \textbf{second} contribution is a demonstration of different strategies to produce diverse Q\&A datasets from a collection of contexts. First, by employing prompt engineering and multi-step LLM querying, then by fine-tuning small LLMs. We compare these strategies to common alternatives based on single prompts to big LLMs. This model can provide an easy-to-use tool to the community for generating diverse RAG Q\&A datasets without expensive queries to big LLMs\footnote{We will make our model public at the time of publication.}.


We believe our proposals contribute a crucial \emph{know-your-data} step to RAG evaluation pipelines, even in cases where private data are involved. It also provides RAG developers with strategies to faithfully evaluate their system's performance by building their own testing datasets. 

\paragraph{Related work:} \citet{GaoEtAl:2024:RAGSurvey} provide a thorough review of strategies for developing a RAG system. Our ideas pertain to their evaluation, and are independent of such development strategies.
Existing evaluation methods \citep{RuEtAl:2024:RagChecker} focus on LLM-assisted metrics for checking aspects like factuality, faithfulness, groundedness, and robustness of generated answers \citep{EsEtAl:2024:Ragas, WuEtAl:2024:ClashEval, KatranidisEtAl:2024:FaaF,ChenEtAl:2024:BenchmarkingLLMsInRAG, LiuEtAl:2023:RECALL, ThakurEtAl:2024:NoMIRACL,AdlakhaEtAl:2024:EvaluatingCorrectnessAndFaithfulness}. Our approach is complementary to these methods as an accurate measurement of these metrics needs a test dataset that is faithful to the type of questions expected in practice. 
Our work also relates to synthetic dataset generation methods \citep{ LongEtAl:2024:LLMDrivenSyntheticDataGenerationSurvey}. Several recent approaches have used LLMs to augment \citep{MøllerEtAl:2024:ParrotDilemma}, label \citep{GilardiEtAl:2023:ChatGPTOutperformsCrowdWorkers,ZiemsEtAl:2024:CanLLMsTransformComputationalSocialScience}, and even generate entirely synthetic datasets \citep{EldanLi:2023:TinyStories}. The RAG dataset generation feature offered by Ragas is closest to us \citep{ragas_synthetic_data}. It uses Evol-Instruct \citep{XuEtAl:2023:WizardLM} to morph simple questions into more complex ones. However, we use a different taxonomy for generating examples and we offer a significantly cheaper fine-tuned generation model. 

\section{Label taxonomy}
\label{section:taxonomy}

Unlike general purpose chatbots, 
enterprise RAG systems have a narrowly defined scope. This allows one to think about the \emph{types} of queries a typical user may ask of the system.  Below we introduce a taxonomy over such types or \emph{labels} that can be used by practitioners across application domains. Our experiments show that this taxonomy is applicable to several commonly used RAG evaluation datasets. If needed, domain experts can also refine it for their specific needs before applying our ideas from the subsequent sections for their analysis.

RAG evaluation datasets generally comprise of (context, query, answer) triplets, where the context (a.k.a. \emph{ground-context}) is expected to contain an answer to the associated query. 
The performance of the retrieval step is based on the system's capability to retrieve the ground-context given a query. Our taxonomy is designed to identify different levels of difficulty for this task. We classify (context, query) pairs based on the \emph{nature} of answer provided by the context to the query. \Cref{table:taxonomy} describes the four classes in our taxonomy - \factsingle{}, \summary{}, \reasoning{}, and \unanswerable{} - along with an example in each case. Classes \factsingle{} and \summary{} require the context to explicitly provide an answer whereas \reasoning{} does not. As the retrieval is done using the contents of the context, it is therefore easier to identify ground-contexts for queries from \factsingle{} and \summary{} classes. We demonstrate these differences in our experiments.

\begin{table*}
\centering
\begin{tabular}{m{1.9cm} m{5.8cm} m{3.2cm} m{3.2cm}}
    \toprule
    \textbf{Class} & \textbf{Description} & \textbf{Example context} & \textbf{Example query} \\
    \midrule
    \factsingle & Answer is present in the context. It has one unit of information and cannot be partially correct. & A table of a sensor's electrical properties & What supply voltage should I use? \\
    \midrule
    \summary & Answer is present in the context. It has multiple units of information. Trading completeness for conciseness yields a partially correct answer. & The conclusion section of a paper & Summarize their key findings for me  \\
    \midrule
    \reasoning & Answer is not explicitly mentioned in the context but can be inferred from it via simple reasoning & An ESG report section on a company's electricity usage & Has there been a net increase in consumption over 5 years? \\
    \midrule
    \unanswerable & Answer is neither present in the context nor can be inferred from it & Claims from a patent on a coffee machine & Is tomato a fruit or a vegetable? \\
    \bottomrule
\end{tabular}
\caption{\label{table:taxonomy}
Proposed taxonomy for classifying (context, query) pairs based on the nature of the request.}
\end{table*}


Queries are not accompanied by a ground-context in practice. However, a RAG developer can likely guess the type of queries expected by the system with respect to the corpus given a narrow enough scope. E.g., a RAG system for referencing specification sheets of electrical sensors would likely get more \factsingle{} queries about properties like input voltage of a sensor.  Similarly, a RAG system that aims to aid an HR professional might be more often used to query procedures and other types of \summary{} information. A system designer would then evaluate their RAG setup on public datasets with an emphasis on its \factsingle{} or \summary{} performance. \citet{YangEtAl:2024:CRAG} proposed a similar taxonomy based on the question alone, while ours looks at a (context, query) pair. The latter bases the label on the type of answer provided by the context to the query. This distinction makes our taxonomy more suitable for evaluating the retrieval step.



\section{Public Datasets}
\label{section:public_data}

We investigate the label composition of Q\&A datasets commonly used for RAG performance evaluation. 
We focus on datasets that contain a well-defined ground-context to help the labelling task and to measure the retrieval performance of the system. 
The datasets considered are: 
\hotpotqa~\cite{YangEtAl:2018:HotPotQA}, 
\msmarco~\cite{NguyenEtAl:2016:MSMARCO}, 
\naturalq~\cite{KwiatkowskiEtAl:2019:NaturalQuestions}, 
\newsqa~\cite{TrischlerEtAl:2017:NewsQA}, 
\pubmedqa~\cite{JinEtAl:2019:PubMedQA}, 
and \squad~\cite{RajpurkarEtAl:2018:SQuAD}.
We use the versions of \hotpotqa~ and \msmarco~ built to train Sentence Transformers~\cite{ReimersGurevych:2019:SBERT}, as they contain the question-answer-ground context triplet needed for this study. Details about data processing and subsampling are mentioned in \Cref{appendix:public_data}.

\section{Labelling examples using LLMs}
\label{section:llm_labeler}

Given the size of typical Q\&A datasets, we turn to LLMs for classification. 
This task typically involves describing all the labels to an LLM and prompting it to select the best match for a given example \citep{EsEtAl:2024:Ragas,ChenEtAl:2024:BenchmarkingLLMsInRAG}. 
We investigate this approach using the prompt detailed in \Cref{appendix:labeller} with two LLMs - \llama{2} \citep{TouvronEtAl:2023:Llama} and \llama{3} \citep{Meta:2024:Llama3}. 
To obtain ground-truth labels, we employed four human annotators to label the same randomly chosen subsets of 100 question-context pairs from each of the six datasets mentioned in \Cref{section:public_data}. 

The quality of the questions analyzed varies significantly: they can be incomplete, ambiguous or completely unintelligible.  
This makes the labelling task difficult and can lead to disagreement between annotators. 
To account for this, we check the level of concordance between the annotators and their majority vote, which is used to define a single label per entry. 
The majority vote discards entries in which a consensus is not found, avoiding ambiguous or otherwise bad quality questions. 
To demonstrate the label variability per annotator, we use Fleiss' kappa \citep{Fleiss:1971:kappa} $\kappa(A_i, M-A_i)$ between an annotator $A_i$ and the majority vote excluding $A_i$ ($M-A_i$). 
We found values of $\kappa(A_i, M-A_i)$ ranging from 0.62 to 0.69, showing a moderate to strong agreement between the annotators and the majority. 
To contrast this behavior to that of an LLMs labeller, we compare the values of $\kappa(\text{LLM}, M-A_i)$ to $\kappa(A_i, M-A_i)$ for a given $A_i$. 
We find that the concordance between $M-A_i$ and \llama{2} is between 67\% to 71\% lower than the concordance between $M-A_i$ and $A_i$. 
\llama{3} performs better, with a concordance only 9\% and 16\% lower than between $M-A_i$ and $A_i$. 
We choose the \llama{3} to generate labels for the following studies, which we will reference as \llamalabels~labels. 

While we observed that \llama{3} tends to correctly differentiate between labels, one noted discrepancy was its preference for \factsingle~over other labels, particularly \summary. 
This confusion is related to how the information requested by the question is present in its ground context: \llama{3} tends to label questions as \factsingle~ even if they ask for multiple pieces of information, if these pieces are contiguous within the context.
For example, given a context that describes a list of devices and their connections, the question ``What devices use a USB cable?'' is a \summary~question because any subset of devices would still be a correct answer, albeit incomplete. \llama{3}, however, classifies this example as \factsingle~if the list of devices is presented as a single sentence. The full confusion matrix is presented in \Cref{appendix:confusion}.

Our LLM-based labelling strategy performs zero-shot classification as the prompt only contains a description of the labels. One could also include (context, query, human label) triplets in the prompt to perform few-shot classification. However, this strategy makes the prompt longer as typical contexts contain several sentences, leading to much higher labelling costs. LLMs also have a finite context window (e.g., 8192 tokens for \llama{3}), which limits the number of triplets that can be included in the prompt, in turn limiting the accuracy of predictions. In principle, one could opt for a higher-cost LLM with a longer context window (e.g., newer versions of \texttt{Llama}), but we restrict ourselves to lower-cost zero-shot classification with \llama{3} in this paper.

\begin{figure}[t]
    \centering
    \includegraphics[width=0.9\linewidth]{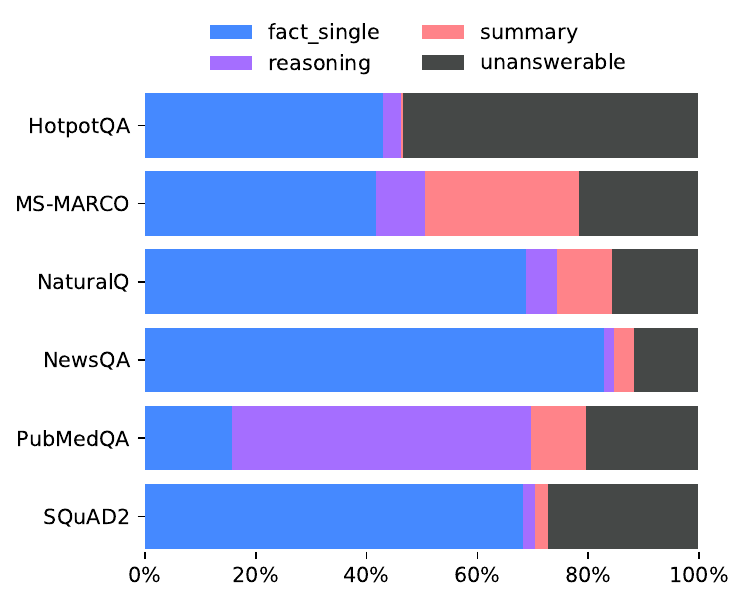}
    \caption{Composition of labels for different datasets.}
    \label{fig:ds_comp}  
\end{figure}


\section{Retrieval performance across classes}
\label{section:experiments}

We now study the performance of the retrieval step of RAG systems as a function of the proposed labels. 
We focus on possible differences when tuning retrieval strategies with different dataset compositions. As a testing setup, we use Elasticsearch~\cite{elasticsearch} to store vector embeddings of the public dataset contexts, which are generated with the \texttt{bge-small-en-v1.5} model ~\cite{bge_embedding}. While dense vectors are highly effective for semantic search, recent applications leverage a hybrid approach, adapting the ranking score with a lexical search component ~\cite{Sawarkar:2024:BlendedRAG}. Elasticsearch provides such a hybrid approach, which can be tuned with a text-weight parameter that varies from 0 (purely vector-based search) to 1 (fully lexical search). Tuning this parameter well is paramount for achieving an optimal performance in a deployed system. However, we show that its optimal value depends not just on the search corpus but also on the types of questions asked.

Recall that Q\&As are associated with a unique ground-context in our problem setup. We characterize the performance of the retriever with the Recall@N metric (for this study, N=5). For each public dataset, we perform retrieval experiments four times: for each label individually (except \unanswerable~questions) and once inclusively for all labels. For each round, we perform text weight scans to find their optimal value. The scan is performed in the following steps: 0, 0.05, 0.1, 0.2, 0.5 and 1. The text-weight steps were chosen empirically, based on the initial results of this investigation. We name the text-weight with the highest Recall@5 the best strategy.

A summary of these experiments is presented in \Cref{table:retrieval_1}, which shows that the best strategy can vary not only across datasets but also across different labels within a dataset. We see relative variations in the best strategy recall from 4.8\% (\naturalq) to 42\% (\newsqa) between best and worst performing labels. The highest recall is achieved with the \factsingle~label most often, while \reasoning~ questions usually achieve the lowest. 
This is expected, as \factsingle~ questions usually contain information directly mentioned in the ground contexts. On the other hand, \reasoning~questions are more difficult to find due to their answers usually being abstractions obtained from their associated contexts. 

More importantly, the best strategy found by using the inclusive dataset, i.e., without any labelling, is not necessarily the same as with individual labels. For example, for \pubmedqa~ the inclusive retrieval prefers a text weight of 0.05 while the \factsingle-only retrieval prefers a dense-vector only search. On the other hand, for \msmarco, the inclusive evaluation would lead to a text weight of 0.1 while the \factsingle-only retrieval optimal text weight is 0.05. We have also tested the hypothesis that the labels influence the text weight choice with different embeddings, such as \texttt{all-MiniLM-L2-v6} \cite{ReimersGurevych:2019:SBERT}, and after applying re-rankers such as \texttt{bge-base} ~\cite{bge_embedding}. These experiments are documented in \Cref{appendix:experiments}. 
These findings show that the performance of RAG systems depends heavily not only on the type of data being searched but also on how the users interact with the system.

\begin{table*}[t]
\centering 
\begin{tabular}{c|c|c|c||c|c}

\textbf{Dataset} & \textbf{Label} & \textbf{Dense} & \textbf{Lexical} & \textbf{Best recall} & \textbf{Best strategy}\\ 

\hline \hline 

\multirow{4}{*}{\hotpotqa} & Inclusive & 0.906 & 0.904 & 0.942 & 0.10\\  \cline{2-6} 
 & \reasoning & 0.890 & 0.878 & \textit{0.924 (-0.076)} & 0.10\\ 
 & \factsingle & 0.891 & 0.897 & 0.930 & 0.10\\ 
 & \summary & 1.000 & 1.000 & \textbf{1.000} & 0.50\\ 

\hline \hline 

\multirow{4}{*}{\msmarco} & Inclusive & 0.752 & 0.719 & 0.804 & 0.10\\  \cline{2-6} 
 & \reasoning & 0.708 & 0.706 & \textit{0.784 (-0.051)} & 0.20\\ 
 & \factsingle & 0.790 & 0.770 & \textbf{0.835} & 0.05\\ 
 & \summary & 0.777 & 0.696 & 0.820 & 0.05\\ 
 
 \hline \hline 
 
\multirow{4}{*}{\naturalq} & Inclusive & 0.686 & 0.464 & \textit{0.686 (-0.033)} & 0.00\\  \cline{2-6} 
 & \reasoning & 0.690 & 0.434 & 0.690 & 0.00\\ 
 & \factsingle & 0.705 & 0.493 & 0.705 & 0.00\\ 
 & \summary & 0.719 & 0.436 & \textbf{0.719} & 0.00\\ 
 
 \hline \hline
 
\multirow{4}{*}{\newsqa} & Inclusive & 0.249 & 0.494 & 0.500 & 0.50\\  \cline{2-6} 
 & \reasoning & 0.194 & 0.379 & \textit{0.379 (-0.161)} & 1.00\\ 
 & \factsingle & 0.262 & 0.533 & \textbf{0.540} & 0.50\\ 
 & \summary & 0.294 & 0.433 & 0.465 & 0.20\\ 
 
 \hline \hline

\multirow{4}{*}{\pubmedqa} & Inclusive & 0.949 & 0.895 & \textit{0.935 (-0.052)} & 0.05\\  \cline{2-6} 
 & \reasoning & 0.947 & 0.885 & 0.947 & 0.00\\ 
 & \factsingle & 0.987 & 0.952 & \textbf{0.987} & 0.00\\ 
 & \summary & 0.985 & 0.959 & 0.985 & 0.00\\ 

\hline \hline

\multirow{4}{*}{\squad} & Inclusive & 0.776 & 0.831 & 0.871 & 0.10\\  \cline{2-6} 
 & \reasoning & 0.757 & 0.671 & \textit{0.789 (-0.104)} & 0.10\\ 
 & \factsingle & 0.818 & 0.852 & \textbf{0.893} & 0.10\\ 
 & \summary & 0.834 & 0.751 & 0.834 & 0.00\\ 
 
\end{tabular} 
\caption{\label{table:retrieval_1}
Summary of retrieval results on different Q\&A labels. Embedding model used is \texttt{bge-small-en-v1.5}. The recall accuracy is measured with Recall@5.} 
\end{table*}

\section{Generating Balanced Datasets}
\label{section:generation}

We now focus on strategies to synthetically generate diverse Q\&A datasets for RAG performance testing.
Several recipes for synthetic dataset generation are found within RAG frameworks, such as LlamaIndex ~\cite{Liu_LlamaIndex_2022} and the RAG Evaluation recipe in the Hugging Face Cookbook ~\cite{hf_cookbook}, which use single prompts to generate question-answer pairs from LLMs. As a benchmark, we generated Q\&A pairs with \llama{3} and the prompt suggested by the latter (also documented in \Cref{appendix:simple_prompt}), on the contexts found in the public datasets described in \Cref{section:public_data}. We utilized the labeling strategy defined in \Cref{section:llm_labeler} and found that 95\% of generated data falls into the \factsingle~ label. As illustrated by the results in \Cref{section:experiments}, this can lead to unrealistic performance expectations when dealing with different types of questions. 

Advanced techniques, such as the ones employed by Ragas~\cite{ragas_synthetic_data}, diversify their generation by sequentially evolving a seed question according to a set of instructions (LLM prompts). While successful in generating datasets with multiple labels, this relies on several LLM queries to generate diverse Q\&As. In addition to being costly, the probability of an LLM hallucination grows with each query. These hallucinations can lead to ``un-grounding'' Q\&As from their original contexts. 
To avoid this, Ragas employs LLM-based critiques at every evolution step to filter out bad examples, which significantly increases the generation cost. 
We choose to ensure this Q\&A-context grounding by inverting the usual generation process: we first build statements based on information from the context and then generate questions that are unambiguously answered by them. 
This strategy reduces the number of LLM queries, grounds the answers, and reduces hallucinations on the question generation by restricting it to a much smaller scope (answer instead of full context).
More information on the Ragas pipeline can be found in \cref{appendix:ragas}.

Our \textbf{statement extraction generation} strategy employs the following steps. \textbf{(1)} The input context is summarized into a sentence (theme). \textbf{(2)} \textbf{Factual statements} are extracted from the context. For completeness, they can include contextualizing information contained in the theme. \textbf{(2.a)} To generate summary questions, we merge the multiple factual statements and the theme into three \textbf{summary statements}. \textbf{(2.b)} To generate reasoning questions, we derive three \textbf{conclusion statements} from the list of factual statements and theme. \textbf{(3)} A random statement is chosen from either the list of factual, summary, or conclusion statements, and a question is generated that is unambiguously answered by it. The theme is once again used to aid with contextualizing information. We used \llama{3} for generation. See \Cref{appendix:statements} for a discussion on different statement strategies and \Cref{appendix:statement_prompts} for the relevant prompts.

\subsection{Model Fine Tuning}

The Q\&A generation strategies previously outlined rely on querying large, state-of-the-art LLMs multiple times. Most methods also include critique steps, in which the quality of the generated dataset is judged, and bad examples are filtered out. This pipeline is costly and can be significantly inefficient if the generated Q\&As are not of good quality. This cost can hinder the performance assessment of RAG systems, particularly for developers with limited access to these large LLMs. To avoid this, we investigate the fine-tuning of small LLMs to generate good quality, diverse Q\&A pairs. To limit consumption at both evaluation and training time, we chose to fine-tune \texttt{Flan-T5-large}~\cite{Chung:2022:t5} with LoRA~\cite{Hu:2021:lora}. Details on the fine-tuning strategy can be found in \Cref{appendix:fine_tuning}.

Six \emph{evaluation} trainings were performed by holding out entries from one specific public dataset at a time. After the models were fine tuned, we generated Q\&As using each held-out public dataset contexts. With this method, we are able to include the impact of generalizability in the model performance by assessing each evaluation training in an independent dataset. The generation step averaged at 15 minutes for generating 2000 Q\&As with a batch size of 64 running on an Apple M1 Max chip. 

\subsection{Synthetic Datasets Quality Comparisons}

We compare the quality of generated Q\&As datasets in the three described setups: with the simple prompt described in the Hugging Face Cookbook, with the statement extraction method, and with the fine-tuned model. 
For the first two cases, \llama{3} is used to generate the questions and build statements. 
As previously stated, 95\% of the generated questions with simple prompt strategy are labelled as \factsingle, therefore we consider this full dataset as being of that type. 

\begin{figure}[t]
  \includegraphics[width=0.98\linewidth]{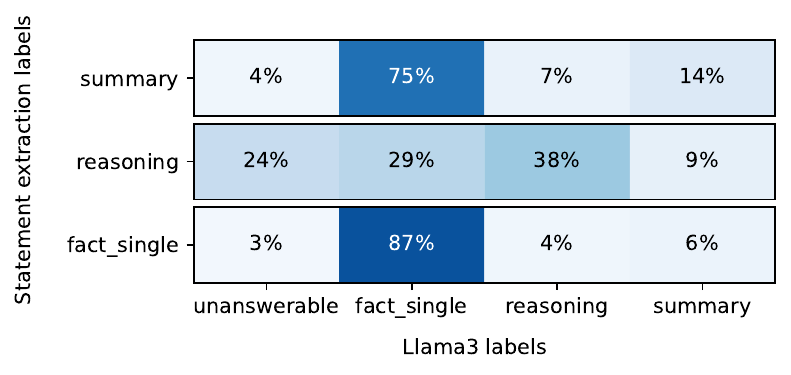} \hfill
  \includegraphics[width=0.98\linewidth]{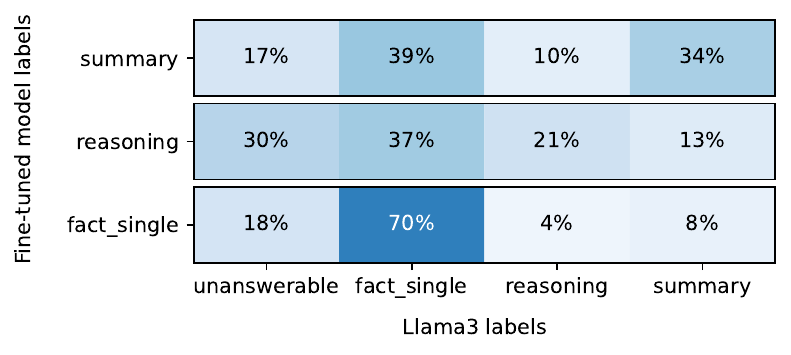} \hfill
  \caption {\label{fig:gen_labels}Distribution of \llamalabels~ labels for statement extraction (top) and fine-tuned model (bottom) per requested label.}
\end{figure}

For the statement extraction method, and the fine-tuned model generation, we find that both are able to produce diverse datasets when prompted with non-\factsingle~labels, as shown in \Cref{fig:gen_labels}.
Alignment between requested label at generation time and \llamalabels~labels is not observed, however, potentially due to two causes: first, as previously stated, \llamalabels~prefers \factsingle~over the other labels due to how the answers are present in the context. Second, not every context equally supports all question types. Some contain mostly factual information statements, for example, the introduction paragraphs of Wikipedia articles. Other contexts can be very small, without enough information to generate independent \reasoning~or \summary~statements. 
In addition, it is important to note that even though the fraction of \unanswerable~questions generated by the fine-tuned model is higher with respect to the statement extraction, the low cost of the former allows users to generate much bigger datasets which can then be cleaned with these labels. 

After selecting Q\&As with valid labels (excluding \unanswerable), we employ LLM-based critiques to further gauge their quality. We choose to apply the following criteria, which are commonly used for this application ~\cite{Liu_LlamaIndex_2022, hf_cookbook}. \textbf{Stand Alone:} whether the question makes sense by itself or if it needs its context to be understood (e.g., questions that mention the word \emph{context} should score low). \textbf{Question Specificity:} how specific the question is to the context (those that are too general, even if answerable by the context, should score low as they are not useful to assess RAG performance). \textbf{Question-Context Grounding:} how well the information requested can be found in the context (questions that cannot be answered should score low). \textbf{Answer-Context Grounding:} how well the information contained in the actual answer can be found in the grounding context. The prompts used to perform these critiques can be found in \Cref{appendix:critique_prompts}. The LLM used to obtain the critiques was \llama{3}.

\begin{figure}[t]
  \includegraphics[width=0.95\linewidth]{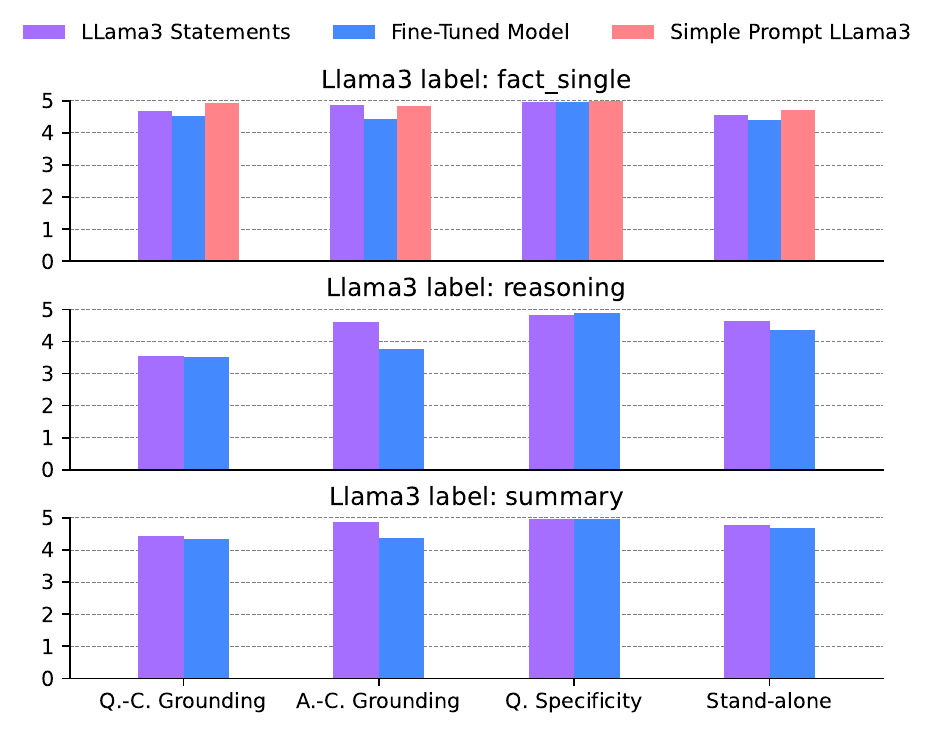} \hfill
  \caption {\label{fig:critiques}Average critique ratings per question label for different Q\&A generation strategies, for all datasets.}
\end{figure}

The critique results are shown in \Cref{fig:critiques}. First, we observe the similarly high scores of both the simple prompt and the statement extraction strategies for \factsingle~ questions. This is consistent with the previous observation that these questions are usually simple statements, containing less information than other labels, thus simpler to label and generate. 
For the other labels, which the simple prompt is unable to generate, we also see generally high ratings for the statement extraction method. Question-context grounding ratings are slightly lower, which we believe is due to the nature of these questions: this critique is more likely to rate the question ``What is the population of Paris?'' (\factsingle) higher than ``What is the role of Paris in EU?'' (\reasoning), even though both are answerable with the first paragraphs of the Paris Wikipedia article, because the first question is partly contained in the text, while the second needs to be inferred.

Finally, we see good agreement between the statement extraction and the fine-tuned model, particularly for \factsingle~and \reasoning. For \summary~questions, the low question-context grounding is consistent with the lower statement extraction rating. Here, the rate of a possible hallucination is similar because, for both cases, the question is generated by an LLM (fine-tuned or \llama{3}). On the other hand, for the answer-context grounding, the statement extraction strategy answer is less likely to be affected by hallucinations because it is based on a statement present in context. While for the fine-tuned model, the LLM needs to construct the answer from the context, conditioned on the question it just generated. We believe the fine-tuned model to be of high value: it is cheaper to generate many examples with it, even if they have to be discarded with LLM-based critiques, than to generate examples with multi-step LLM querying that also need to be filtered.

\section{Conclusion}

In this study, we present tools to build synthetic datasets aimed at evaluating RAG systems and strategies to characterize these datasets in terms of information request labels. 
We show that public Q\&A datasets, and synthetic datasets generated with simple LLM prompts, can be highly unbalanced in terms of these labels, and that the retrieval performance of common RAG strategies depend on them. The combination of these two observations can lead to non-optimal design choices when building a RAG system if the type of user interactions is not reflected in the evaluation dataset. 
To mitigate this issue, we present strategies to generate diverse synthetic data. 
First, we propose a statement extraction strategy to generate grounded and labelled Q\&As, and then we fine-tune a small LLM to perform the Q\&A generation. 
Both strategies are successful in generating high quality, diverse Q\&A datasets. 
While these strategies still require a second step of quality evaluation and cleaning, we believe they are more efficient in terms of cost and performance than current available solutions. 
These proposals constitute an important step in empowering RAG developers to properly evaluate and optimize their own systems.
Even though our study focuses on the impact of the labeling strategy on the retrieval performance, further experiments on the response generation step may also be of interest.

\bibliography{custom}
\newpage
\textcolor{white}{.}
\newpage

\appendix

\section{Public Datasets}
\label{appendix:public_data}

\Cref{table:data_all} lists the public datasets we use. The pre-processing for each dataset is described below. In each case, we only consider contexts with at most 10 000 characters.

\begin{table*}[t]

    \begin{tabular*}{\linewidth}{@{\extracolsep{\fill}} lrrrrrr}
        \toprule
\textbf{Dataset} & \textbf{Contexts} & \textbf{Label Q\&As} & \textbf{fact\_single} & \textbf{reasoning} & \textbf{summary} & \textbf{unanswerable} \\
        \midrule
        
\textbf{HotpotQA} & 65 489 & 5000 & 42.9\% & 3.4\% & 0.2\% & \textbf{53.4\%}\\ 
\textbf{MS MARCO} & 808 712 & 4972 & \textbf{41.8\%} & 8.8\% & 27.9\% & 21.5\%\\ 
\textbf{NaturalQ} & 111 388 & 5000 & \textbf{68.9\%} & 5.6\% & 9.9\% & 15.6\%\\ 
\textbf{NewsQA} & 89 481 & 6890 & \textbf{83.0\%} & 1.8\% & 3.6\% & 11.7\%\\ 
\textbf{PubMedQA} & 61 243 & 6847 & 15.8\% & \textbf{53.9\%} & 9.9\% & 20.3\%\\ 
\textbf{SQuAD2} & 19 029 & 6910 & \textbf{68.2\%} & 2.2\% & 2.4\% & 27.1\%\\

        \bottomrule
    \end{tabular*}
    \caption{\label{table:data_all}
    Datasets considered in the study.}
\end{table*}

\paragraph{SQuAD2:} We use the training set of SQuAD2 \citep{RajpurkarEtAl:2018:SQuAD}. In this dataset, question-answer pairs are associated with specific paragraphs from Wikipedia articles. We treat each paragraph as a separate context, resulting in 19 029 unique contexts. We merge the question-answer pairs from all paragraphs into a list and randomly sample 6910 pairs along with their associated contexts for labeling.

\paragraph{NewsQA:} This dataset consists of QA pairs, each associated with a news story \citep{TrischlerEtAl:2017:NewsQA}. We use entire news stories as individual contexts and randomly sample 6890 context-question-answer triplets for labeling.

\paragraph{PubMedQA:} We use the unlabelled subset of this dataset that has 61 243 eligible contexts, each corresponding to the abstract of a research article \citep{JinEtAl:2019:PubMedQA}. Each context is accompanied by a question and an answer. We retain a subset of 68 47 randomly sampled examples for labeling.

\paragraph{HotpotQA:} The original HotpotQA dataset was designed to test the ability of QA systems to iteratively combine information across multiple contexts \citep{YangEtAl:2018:HotPotQA}. Consequently, each question in this dataset is associated with several contexts. While this is an interesting use case, we are primarily concerned with the one context, one question setting. Therefore, we use the version of the dataset used in \citet{ReimersGurevych:2019:SBERT} for training a sentence similarity model. In this version, each question is associated with a relevant \emph{positive} context and an irrelevant \emph{negative} context. We randomly sampled 5000 of the available 65 489 (question, positive context) pairs for labeling. Due to the nature of the dataset, most of the sampled questions cannot be answered using the single \emph{positive} context alone. This is evident from \Cref{table:data_all}, which shows that our labeller marks a majority of the HotpotQA questions as unanswerable.

\paragraph{MS MARCO:} We obtained v2.1 of this dataset from Hugging Face \citep{NguyenEtAl:2016:MSMARCO}. Each question has 10 passages (top-10 hits on Bing) associated with it. We concatenated these passages to get one context per question. We then randomly selected 5000 examples. Of these, 28 contained characters that \llama{3} was unable to process. This left us with 4972 labelled (context, question, answer) triplets.

\paragraph{NaturalQ:} We obtained the simplified training set of the natural questions dataset \citep{KwiatkowskiEtAl:2019:NaturalQuestions}. Each row contains a question along with a \emph{long answer} comprising paragraphs from Wikipedia that contain an answer to the question. We concatenate all long answers associated with a question to get the corresponding context and then randomly sample 5000 out of 111 388 examples for labeling.

\section{\llama{3} confusion matrix}
\label{appendix:confusion}

\begin{figure}[t]
    \centering
    \includegraphics[width=0.98\linewidth]{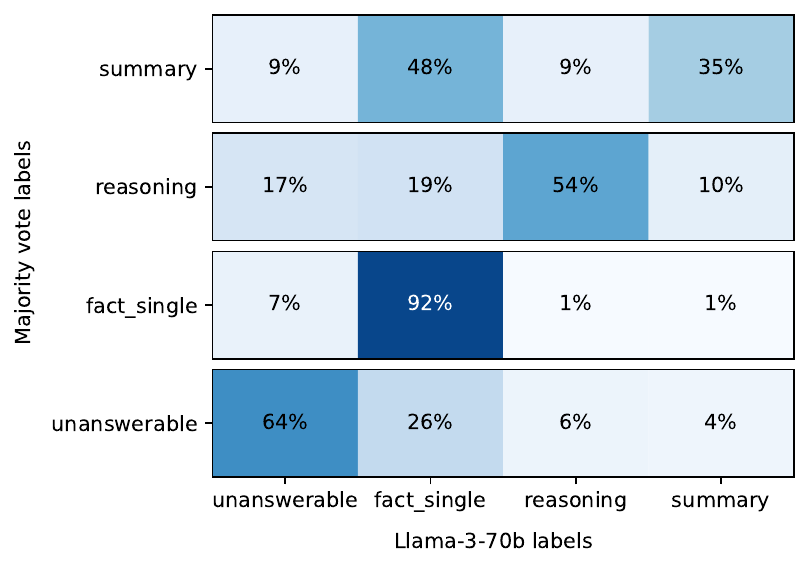}
    \caption{Confusion matrix between labels given by \llama{3} and the annotators' majority vote. Each row shows the distribution of \llama{3} labels given a majority vote label.}
    \label{fig:confusion}  
\end{figure}

\Cref{fig:confusion} compares the labels assigned by \llama{3} to the labels selected by a majority of the human annotators.
\section{Retrieval Experiments}
\label{appendix:experiments}


\begin{table*}[t]
\centering 
\begin{tabular}{c|c|c|c||c|c}

\textbf{Dataset} & \textbf{Label} & \textbf{Dense} & \textbf{Lexical} & \textbf{Best recall} & \textbf{Best strategy}\\ 

\hline \hline

\multirow{4}{*}{\hotpotqa} & Inclusive & 0.933 & 0.946 & 0.965 & 0.10\\  \cline{2-6} 
 & \reasoning & 0.907 & 0.919 & \textit{0.948 (-0.052)} & 0.10\\ 
 & \factsingle & 0.926 & 0.938 & 0.954 & 0.10\\ 
 & \summary & 1.000 & 1.000 & \textbf{1.000} & 0.10\\ 

\hline \hline 

\multirow{4}{*}{\msmarco} & Inclusive & 0.790 & 0.780 & 0.795 & 0.50\\  \cline{2-6} 
 & \reasoning & 0.749 & 0.763 & \textit{0.749 (-0.093)} & 0.00\\ 
 & \factsingle & 0.825 & 0.822 & \textbf{0.842} & 0.20\\ 
 & \summary & 0.822 & 0.788 & 0.822 & 0.00\\ 
 
 \hline \hline 
 
\multirow{4}{*}{\naturalq} & Inclusive & 0.695 & 0.580 & 0.695 & 0.00\\  \cline{2-6} 
 & \reasoning & 0.673 & 0.562 & \textit{0.687 (-0.056)} & 0.10\\ 
 & \factsingle & 0.723 & 0.614 & \textbf{0.743} & 0.05\\ 
 & \summary & 0.715 & 0.560 & 0.715 & 0.00\\ 
 
 \hline \hline
 
\multirow{4}{*}{\newsqa} & Inclusive & 0.300 & 0.461 & 0.463 & 0.50\\  \cline{2-6} 
 & \reasoning & 0.210 & 0.315 & \textit{0.323 (-0.178)} & 0.20\\ 
 & \factsingle & 0.320 & 0.498 & \textbf{0.501} & 0.50\\ 
 & \summary & 0.302 & 0.384 & 0.388 & 0.50\\ 
 
 \hline \hline

\multirow{4}{*}{\pubmedqa} & Inclusive & 0.892 & 0.908 & 0.896 & 0.05\\  \cline{2-6} 
 & \reasoning & 0.887 & 0.907 & \textit{0.887 (-0.078)} & 0.00\\ 
 & \factsingle & 0.941 & 0.961 & 0.965 & 0.50\\ 
 & \summary & 0.965 & 0.977 & \textbf{0.965} & 0.00\\ 

\hline \hline

\multirow{4}{*}{\squad} & Inclusive & 0.659 & 0.828 & 0.830 & 0.50\\  \cline{2-6} 
 & \reasoning & 0.618 & 0.757 & \textit{0.763 (-0.106)} & 0.20\\ 
 & \factsingle & 0.717 & 0.867 & \textbf{0.869} & 0.50\\ 
 & \summary & 0.716 & 0.834 & 0.852 & 0.20\\ 
 
\end{tabular} 
\caption{\label{table:retrieval_2}
Embedding model: \texttt{bge-small-en-v1.5}; Re-ranking model: \texttt{BGE-base}} 
\end{table*}


\begin{table*}[t]
\centering 
\begin{tabular}{c|c|c|c||c|c}

\textbf{Dataset} & \textbf{Label} & \textbf{Dense} & \textbf{Lexical} & \textbf{Best recall} & \textbf{Best strategy}\\ 

\hline \hline

\multirow{4}{*}{\hotpotqa} & Inclusive & 0.830 & 0.904 & 0.929 & 0.10\\  \cline{2-6} 
 & \reasoning & 0.767 & 0.878 & \textit{0.907 (-0.093)} & 0.10\\ 
 & \factsingle & 0.813 & 0.897 & 0.914 & 0.10\\ 
 & \summary & 0.818 & 1.000 & \textbf{1.000} & 0.10\\ 

\hline \hline 

\multirow{4}{*}{\msmarco} & Inclusive & 0.697 & 0.719 & 0.799 & 0.10\\  \cline{2-6} 
 & \reasoning & 0.661 & 0.706 & \textit{0.781 (-0.053)} & 0.20\\ 
 & \factsingle & 0.740 & 0.770 & \textbf{0.834} & 0.10\\ 
 & \summary & 0.711 & 0.696 & 0.801 & 0.05\\ 
 
 \hline \hline 
 
\multirow{4}{*}{\naturalq} & Inclusive & 0.625 & 0.464 & 0.625 & 0.00\\  \cline{2-6} 
 & \reasoning & 0.623 & 0.434 & \textit{0.623 (-0.018)} & 0.00\\ 
 & \factsingle & 0.641 & 0.493 & \textbf{0.641} & 0.00\\ 
 & \summary & 0.640 & 0.436 & 0.640 & 0.00\\ 
 
 \hline \hline
 
\multirow{4}{*}{\newsqa} & Inclusive & 0.186 & 0.494 & 0.496 & 0.50\\  \cline{2-6} 
 & \reasoning & 0.177 & 0.379 & \textit{0.379 (-0.156)} & 1.00\\ 
 & \factsingle & 0.195 & 0.533 & \textbf{0.535} & 0.50\\ 
 & \summary & 0.229 & 0.433 & 0.441 & 0.50\\ 
 
 \hline \hline

\multirow{4}{*}{\pubmedqa} & Inclusive & 0.886 & 0.895 & \textit{0.902 (-0.075)} & 0.50\\  \cline{2-6} 
 & \reasoning & 0.877 & 0.885 & 0.922 & 0.10\\ 
 & \factsingle & 0.944 & 0.952 & 0.969 & 0.05\\ 
 & \summary & 0.935 & 0.959 & \textbf{0.977} & 0.10\\ 

\hline \hline

\multirow{4}{*}{\squad} & Inclusive & 0.773 & 0.831 & 0.878 & 0.10\\  \cline{2-6} 
 & \reasoning & 0.743 & 0.671 & \textit{0.803 (-0.096)} & 0.10\\ 
 & \factsingle & 0.816 & 0.852 & \textbf{0.899} & 0.10\\ 
 & \summary & 0.852 & 0.751 & 0.852 & 0.05\\ 
 
\end{tabular} 
\caption{\label{table:retrieval_3}
Embedding model: \texttt{all-minilm-l6-v2}; Rerank: False} 
\end{table*}


\begin{table*}[t]
\centering 
\begin{tabular}{c|c|c|c||c|c}

\textbf{Dataset} & \textbf{Label} & \textbf{Dense} & \textbf{Lexical} & \textbf{Best recall} & \textbf{Best strategy}\\ 

\hline \hline

\multirow{4}{*}{\hotpotqa} & Inclusive & 0.881 & 0.946 & 0.959 & 0.10\\  \cline{2-6} 
 & \reasoning & 0.814 & 0.919 & \textit{0.814 (-0.186)} & 0.00\\ 
 & \factsingle & 0.868 & 0.938 & 0.868 & 0.00\\ 
 & \summary & 1.000 & 1.000 & \textbf{1.000} & 0.50\\ 

\hline \hline 

\multirow{4}{*}{\msmarco} & Inclusive & 0.773 & 0.780 & 0.804 & 0.20\\  \cline{2-6} 
 & \reasoning & 0.724 & 0.763 & \textit{0.784 (-0.051)} & 0.50\\ 
 & \factsingle & 0.814 & 0.822 & 0.814 & 0.00\\ 
 & \summary & 0.810 & 0.788 & \textbf{0.835} & 0.20\\ 
 
 \hline \hline 
 
\multirow{4}{*}{\naturalq} & Inclusive & 0.668 & 0.580 & 0.668 & 0.00\\  \cline{2-6} 
 & \reasoning & 0.630 & 0.562 & 0.658 & 0.05\\ 
 & \factsingle & 0.697 & 0.614 & \textbf{0.722} & 0.05\\ 
 & \summary & 0.655 & 0.560 & \textit{0.655 (-0.067)} & 0.00\\ 
 
 \hline \hline
 
\multirow{4}{*}{\newsqa} & Inclusive & 0.239 & 0.461 & 0.462 & 0.50\\  \cline{2-6} 
 & \reasoning & 0.218 & 0.315 & \textit{0.331 (-0.169)} & 0.10\\ 
 & \factsingle & 0.254 & 0.498 & \textbf{0.500} & 0.50\\ 
 & \summary & 0.229 & 0.384 & 0.388 & 0.10\\ 
 
 \hline \hline

\multirow{4}{*}{\pubmedqa} & Inclusive & 0.878 & 0.908 & \textit{0.899 (-0.078)} & 0.05\\  \cline{2-6} 
 & \reasoning & 0.869 & 0.907 & 0.913 & 0.20\\ 
 & \factsingle & 0.938 & 0.961 & 0.938 & 0.00\\ 
 & \summary & 0.956 & 0.977 & \textbf{0.977} & 1.00\\ 

\hline \hline

\multirow{4}{*}{\squad} & Inclusive & 0.678 & 0.828 & 0.830 & 0.50\\  \cline{2-6} 
 & \reasoning & 0.625 & 0.757 & 0.757 & 0.50\\ 
 & \factsingle & 0.738 & 0.867 & \textit{0.738 (-0.108)} & 0.00\\ 
 & \summary & 0.722 & 0.834 & \textbf{0.846} & 0.10\\ 
 
\end{tabular} 
\caption{\label{table:retrieval_4}
Embedding model: \texttt{all-minilm-l6-v2}; Re-ranking model: \texttt{BGE-base}} 
\end{table*}


\Cref{table:retrieval_1,,table:retrieval_2,,table:retrieval_3,,table:retrieval_4} show the dependence of the retrieval performance on proposed taxonomy. We quantify the results with the use of the Recall@5 metric. \Cref{table:retrieval_1,,table:retrieval_2} employ the embedding model \texttt{bge-small-en-v1.5}, with the latter also re-ranking the retrieval results with the \texttt{bge-base} re-ranker. \Cref{table:retrieval_3,,table:retrieval_4} employ the embedding model \texttt{all-minilm-l6-v2}, with the latter also re-ranking the retrieval results with the \texttt{bge-base} re-ranker. 
We do not show the results for the \unanswerable~label as that is not a label of interest for our study. However, these questions are included in the ``Inclusive'' evaluation, which reflects the standard use of these datasets by the public. For this reason, the retrieval performance of the ``Inclusive'' evaluation can be lower than any of the displayed individual labels. The results for the \hotpotqa~ \summary~label are statistically limited by the number of Q\&As present in the analyzed dataset after labelling -- we leave them in the results tables for completeness.

\section{Labelling Prompt}
\label{appendix:labeller}

\begin{quote}
\begin{lstlisting}
Consider the following context information and a related question.
-- Context start --

[[{context}]]

-- Context end --

-- Question start --

[[{question}]]

-- Question end --

Select the most suitable label from the list below:

{label_name: fact_single, label_description: A complete answer to this question is explicitly mentioned in the context and is a single simple value}
{label_name: summary, label_description: A complete answer to this question is explicitly mentioned in the context and is more like a summary, a procedure for doing something, or a composite of multiple parts}
{label_name: reasoning, label_description: A complete answer to this question is not explicitly mentioned in the context but can be inferred from the information given in it}
{label_name: unanswerable, label_description: A complete answer to this question is neither explicitly mentioned in the context nor can be inferred from the information given in it}

Return your response in the following JSON format: {"label_name": "selected_label_name", "reason": "reason_for_your_choice"}

You must select exactly one label from the list above. Do not select anything that is not in the list. Do not return anything other than the JSON format requested above.
\end{lstlisting}
\end{quote}
\section{Simple Prompt}
\label{appendix:simple_prompt}

The simple prompt used in this study was obtained from the Hugging Face RAG Evaluation Cookbook \cite{hf_cookbook} with a small modification to generate a python dictionary. We found this generation style worked well with \llama{3} and avoided missing questions and/or answers.

\begin{quote}
\begin{lstlisting}
Your task is to write a factoid question and an answer given a context.
Your factoid question should be answerable with a specific, concise piece of factual information from the context.
Your factoid question should be formulated in the same style as questions users could ask in a search engine.
This means that your factoid question MUST NOT mention something like "according to the passage" or "context".

Provide your answer as a JSON dictionary as follows:

Output:::
{{"question": "your factoid question", "answer": "your answer to the factoid question"}}

Now here is the context.

Context: 
[[{context}]]
Output:::
\end{lstlisting}
\end{quote}

\section{Discussion on Statements}
\label{appendix:statements}

The final generated question depends on how its source statement, i.e., answer, was generated. Factual statements focus on unitary pieces of factual information directly contained in the context. For example, parsing the first couple of sentences from the Wikipedia article on Paris, the generated factual statements would be such as ``Paris is the capital of France'' (which would answer the question ``What is the capital of France?''), ``The population of Paris is estimated to be 2,102,650 residents as of January 2023'' (``What is the population of Paris?''), ``The Paris Region had a GDP of 765 billion euros in 2021.'' (``What is the GDP of Paris?''), etc. To generate a summary statement, information is combined into composite sentences. In the previous example, a summary statement would be ``Paris, the capital and largest city of France, has a population of approximately 2.1 million residents as of 2023.'' (which would answer the composite question ``What is the capital of France and what is its population?'' or the indirect question ``What is the population of the capital of France?''). For conclusion statements, we ask the LLM to infer statements that are not directly included in the original factual statements list. In this case, one possible conclusion statement would be ``Paris is a significant economic hub in the European Union, given its large population and high GDP.'', which answers the question ``What is the role of Paris in the European Union?''.

The usage of themes to ground both the extracted statements and question generation comes from the observed difference between \emph{corpus-level} questions and \emph{document-level} questions. This differentiation is related to a broad categorization of RAG applications as corpus-level or document-level. Corpus-level RAG involves multiple documents which can include multiple themes, while document-level RAG generally contains a narrower scope. In the previous example, we could expect users to ask questions in a different manner if performing RAG over the entire Wikipedia collection of articles, as opposed to directly querying the article on Paris. In the former case, the user would more likely craft a more specific question (``What is the population \textit{of Paris}?''), while in the latter, we can expect less specification (``What's the city's population?''). We observed that the usage of themes favored the more specific corpus-level questions, while omitting it led to less specific document-level questions.
\section{Statement Extraction Prompts}
\label{appendix:statement_prompts}

\subsection{Theme}

\begin{quote}
\begin{lstlisting}
In a few words, extract the main theme behind the following passage: [[{context}]]
\end{lstlisting}
\end{quote}

\subsection{Factual statements}

\begin{quote}
\begin{lstlisting}
Extract at most five factual statements based on the following passage and its theme. You need to strictly comply with the following guidelines:
- Each statement must contain a single unit of factual information.
- Each statement must be written in the style of an answer to a factual question. 
- Each statement must be understandable without the aid of any other source of information.
- Each statement must include contextual information derived from the passage theme.
- Each statement must only contain information that exists in the original passage and theme.
- Each statement must be independent from the other statements.

Generate the statements as a bullet list with the following format:
> Statement
> Statement
etc

Theme: [[{theme}]]
Passage: [[{context}]]
\end{lstlisting}
\end{quote}

\subsection{Summary statements}

\begin{quote}
\begin{lstlisting}
Merge the following sentences into three summary statements. 
Each summary statement must summarise information contained in more than one sentence. 
Each summary statement must be independent and non-overlapping. 
Each summary statement should be a complete sentence. 
Each summary statement can include contextual information contained in the theme below. 
Each summary statement must be understandable without the aid of any other source of information.

Generate the statements as a bullet list with the following format:
> Summary statement
> Summary statement
> Summary statement

Theme: [[{theme}]]

Sentences:[[
{statements}
]]
\end{lstlisting}
\end{quote}

\subsection{Reasoning statements}

\begin{quote}
\begin{lstlisting}
Generate three reasoning conclusions that can be drawn from the following statements. 
A reasoning conclusion is an inferred piece of information obtained from critically analysing a group of multiple statements. 
Reasoning conclusions do not contain information directly contained on any statements. 
Each conclusion must be independent and non-overlapping. 
Each conclusion should be a complete sentence. 
Each conclusion must be understandable without the aid of any other source of information.
Each conclusion can include contextual information contained in the theme below. 

Generate the conclusions as a bullet list with the following format:
> conclusion
> conclusion
> conclusion
etc

Theme: [[{theme}]]

Statements:[[
{statements}
]]
\end{lstlisting}
\end{quote}

\subsection{Question}

\begin{quote}
\begin{lstlisting}
I have a paragraph with the following theme:
[[{theme}]]

From this paragraph, I extracted the following statement:
[[{statement}]]

Generate one question which is answered only by the statement above. 
In order to avoid generic questions, use contextual information from the theme to formulate the question. 
The question should be concise and in the style of a user asking questions to a search engine.

Generate the question as a bullet list with the following format:
> Question
Do not output anything else other than the question.
\end{lstlisting}
\end{quote}
\section{Model Fine-Tuning}
\label{appendix:fine_tuning}

Our fine-tuning strategy starts with the FLAN-T5 family of models ~\cite{Chung:2022:t5}. We found that the \texttt{small} and \texttt{base} model sizes were not perceptive enough to extract interesting information from the contexts used, with the \texttt{large} model size being the smallest model that achieved that goal. Keeping in mind the objective of providing a low-resources strategy, we employ LoRA ~\cite{Hu:2021:lora} in the fine-tuning step, training 30\% of the 785M parameters in the \texttt{Flan-T5-large} model. 

The fine-tuning training data contains the contexts extracted from the public datasets described in \Cref{section:public_data} as inputs. The outputs are the Q\&As generated through the answer-first statement extraction method described previously. The final dataset contained 2000\footnote{The number of training examples is limited by the generation of Q\&As with the standard, multi-step methods.} context-Q\&As per type, per public dataset to a total of 36k entries, from which 20\% was held out for validation.  

In order to allow for the generation of multiple question types with the same fine-tuned model, we add a question type flag (\texttt{\textlangle\textlangle fact\_single\textrangle\textrangle}, \texttt{\textlangle\textlangle summary\textrangle\textrangle} or \texttt{\textlangle\textlangle reasoning\textrangle\textrangle}) to the beginning of each context to identify which Q\&A type will be used as target. The Q\&A is represented by a single string separated by the token ``\textlangle a\textrangle'', which is added to the T5 model tokenizer. Therefore, the fine-tuning step sees each context three times, each time with a different question type flag and a different associated Q\&A. In summary, the inputs and outputs used for the fine-tuning are as follows.

\begin{quote}
\begin{lstlisting}
Input: <<question_type>> Ground truth context
Output: <<question_type>> Statement extraction question <a> Statement extraction answer
\end{lstlisting} 
\end{quote}
\section{Critique Prompts}
\label{appendix:critique_prompts}

The prompts described here are adapted from \cite{hf_cookbook}. The ratings obtained range from 1 to 5. For visualization purposes, they are scaled to range from 0 to 5.

\begin{quote}
\begin{lstlisting}
In a few words, extract the main theme behind the following passage: [[{context}]]
\end{lstlisting}
\end{quote}

\begin{quote}
\begin{lstlisting}
q_to_c_groundedness:
You will be given a context and a sentence that should be a question.
Your task is to provide a 'total rating' scoring how well one can answer the given question unambiguously with the given context.
Give your answer on a scale of 1 to 5, where 1 means that the question is not answerable at all given the context, and 5 means that the question is clearly and unambiguously answerable with the context. 
If the sentence provided is not actually a question, rate it as 1.

Provide your answer as a python dictionary as follows:

Answer:::
{{"evaluation": "Your rationale for the rating, as a brief and concise text", "rating": "your rating, as a number between 1 and 5"}}

You MUST provide values for 'evaluation' and 'rating' in your answer. Provide ONLY the python dictionary as your answer.

Now here are the question and context.

Question: "{question}"

Context: "{context}"

Answer:::
\end{lstlisting}
\end{quote}

\begin{quote}
\begin{lstlisting}
a_to_c_groundedness:
You will be given a context, and a passage. 
Your task is to provide a 'total rating' scoring how well the statements in the provided passage can be infered from the provided context. 
Give your rating on a scale of 1 to 5, where 1 means that none of the statements in the passage can be inferred from the provided context, while 5 means that all of the statements in the passage can be unambiguously and entirely obtained from the context.

Provide your answer as a python dictionary as follows:

Answer:::
{{"evaluation": "Your rationale for the rating, as a brief and concise text", "rating": "your rating, as a number between 1 and 5"}}

You MUST provide values for 'evaluation' and 'rating' in your answer. Provide ONLY the python dictionary as your answer.

Now here are the context and statement.

Context: "{context}"

Passage: "{answer}"

Answer:::
\end{lstlisting}
\end{quote}

\begin{quote}
\begin{lstlisting}
q_feasibility:
You will be given a context and a question.
This context is extracted from a collection of passages, and the question will be used to find it. 
Your task is to provide a 'total rating' scoring how well this context can be retrieved based on the specificity and pertinence of the question.
Give your answer on a scale of 1 to 5, where 1 means that it will be difficult to find this context from this question due to lack of specificity or pertinence, and 5 means that the context can clearly be found with information contained in the question. 

Provide your answer as a python dictionary as follows:

Answer:::
{{"evaluation": "Your rationale for the rating, as a brief and concise text", "rating": "your rating, as a number between 1 and 5"}}

You MUST provide values for 'evaluation' and 'rating' in your answer. Provide ONLY the python dictionary as your answer.

Now here are the question and context.

Question: "{question}"

Context: "{context}"

Answer:::
\end{lstlisting}
\end{quote}

\begin{quote}
\begin{lstlisting}
stand_alone:
You will be given a question.
Your task is to provide a 'total rating' representing how context-independent this question is.
Give your answer on a scale of 1 to 5, where 1 means that the question depends on additional information to be understood, and 5 means that the question makes sense by itself.
For instance, if the question refers to a particular setting, like 'in the context' or 'in the document', the rating must be 1.
The questions can contain obscure technical nouns or acronyms and still be a 5: it must simply be clear to an operator with access to documentation what the question is about.

For instance, "What is the name of the checkpoint from which the ViT model is imported?" should receive a 1, since there is an implicit mention of a context, thus the question is not independent from the context.

Provide your answer as a python dictionary as follows:

Answer:::
{{"evaluation": "Your rationale for the rating, as a brief and concise text", "rating": "your rating, as a number between 1 and 5"}}

You MUST provide values for 'evaluation' and 'rating' in your answer. Provide ONLY the python dictionary as your answer.

Now here is the question.

Question: "{question}"

Answer:::
\end{lstlisting}
\end{quote}

\begin{quote}
\begin{lstlisting}
q_usefulness:
You will be given a question. 
This question is to be used to find information in a collection of documents. 
Your task is to provide a 'total rating' representing how useful this question can be to a user with domain knowledge on the subject covered by the document collection. 
Give your answer on a scale of 1 to 5, where 1 means that the question is not useful at all, and 5 means that the question is extremely useful.

Provide your answer as a python dictionary as follows:

Answer:::
{{"evaluation": "Your rationale for the rating, as a brief and concise text", "rating": "your rating, as a number between 1 and 5"}}

You MUST provide values for 'evaluation' and 'rating' in your answer. Provide ONLY the python dictionary as your answer.

Now here is the question.

Question: "{question}"

Answer:::
\end{lstlisting}
\end{quote}

\begin{quote}
\begin{lstlisting}
c_usefulness:
You will be given a context. 
This context is a part of a collection of contexts that users can query. 
Your task is to provide a 'total rating' representing how useful this context can be to extract statements for a user with domain knowledge on the subject covered by the context collection. 
Give your answer on a scale of 1 to 5, where 1 means that the context does not contain any useful statements, and 5 means that the context contains multiple statements that provide the user with different pieces of information.

Provide your answer as a python dictionary as follows:

Answer:::
{{"evaluation": "Your rationale for the rating, as a brief and concise text", "rating": "your rating, as a number between 1 and 5"}}

You MUST provide values for 'evaluation' and 'rating' in your answer. Provide ONLY the python dictionary as your answer.

Now here is the context.

Context::: "{context}"

Answer:::
\end{lstlisting}
\end{quote}

\begin{quote}
\begin{lstlisting}
c_clarity:
You will be given a context. 
This context is a part of a collection of contexts that users can query. 
Your task is to provide a 'total rating' representing the clarity of the information contained in the context. 
Give your answer on a scale of 1 to 5, where 1 means that the context contains incomplete, unclear or poorly formatted information, and 5 means that the context contains only complete, clear and well formatted statements. 

Provide your answer as a python dictionary as follows:

Answer:::
{{"evaluation": "Your rationale for the rating, as a brief and concise text", "rating": "your rating, as a number between 1 and 5"}}

You MUST provide values for 'evaluation' and 'rating' in your answer. Provide ONLY the python dictionary as your answer.

Now here is the context.

Context::: "{context}"

Answer:::
\end{lstlisting}
\end{quote}

\begin{quote}
\begin{lstlisting}
qa_tautology:
You will be given a question and passage its answer.
Your question is to judge whether this question and answer pair form a tautological exchange.
Give your answer on a scale of 1 to 5, where 1 means that the question and answer repeat the same information, and 5 means that the answer is made of entirely new information.

Provide your output as a python dictionary as follows:

Output:::
{{"evaluation": "Your rationale for the rating, as a brief and concise text", "rating": "your rating, as a number between 1 and 5"}}

You MUST provide values for 'evaluation' and 'rating' in your answer. Provide ONLY the python dictionary as your answer.

Now here are the question and its answer.

Question::: "{question}"

Answer::: "{answer}"

Output:::
\end{lstlisting}
\end{quote}
\section{Dataset generation using Ragas}
\label{appendix:ragas}

We discussed our approach for generating RAG evaluation datasets in \Cref{section:generation}. Ragas offers a similar feature \citep{ragas_synthetic_data} based on the Evol-Instruct framework \citep{XuEtAl:2023:WizardLM}. Evol-Instruct was originally developed to generate complex questions by \emph{evolving} a set of simpler \emph{seed} questions. For example, starting with the seed question ``what is the boiling point of water?'' the so called \emph{add-constraint} evolution asks an LLM to make the question more complex by adding a constraint to it. This, for instance, would lead to an output like ``what is the boiling point of water at 5 atm pressure?'' Evol-Instruct defines many such evolution prompts. Ragas adapts three of them to the RAG setting - \emph{simple}, \emph{multi-context} and \emph{reasoning}.

Ragas begins by generating a seed question that can be answered by the given context. There are no additional requirements on the type of this seed question. The \emph{simple} evolution simply returns this seed question. \emph{Multi-context} combines two contexts and generates a question that can only be answered by reading both contexts. The \emph{reasoning} evolution is similar to our reasoning class in \Cref{table:taxonomy}, and requires a logical chain of reasoning to infer an answer to the question. Users can specify the relative proportion of these three evolutions in the generated dataset.

Two differences between our taxonomy in \Cref{table:taxonomy} and Ragas' evolutions are immediately obvious. First, Ragas lacks a counterpart for our \summary{} class. One might assume that the \emph{multi-context} evolution is similar to \summary{}. However, this is not true as \emph{multi-context} evolution only requires the answer to combine information from multiple contexts. This answer need not contain multiple facts, as required by \summary{}. Second, the \emph{simple} evolution is not a pure class as per our taxonomy. This evolution just returns the seed question, which was generated without any requirement on the expected answer type. Owing to these differences, it is not possible to directly use Ragas to generate questions according to our taxonomy.

We generated 600 (context, question, answer) triplets for each public dataset in \Cref{table:data_all} using Ragas. In each case, we used the \emph{simple} and \emph{reasoning} evolutions in equal proportion. The \emph{multi-context} evolution associates more than one context per question as explained above, and hence is outside our scope. Our first attempt at generating these examples using \llama{3} failed as a significant number of questions returned by Ragas included part of the question-generation prompt used by the library. We then switched to \llama{2} but, despite multiple attempts, this lead to unresolved \texttt{AssertionError} while generating the dataset. Eventually, we were able to successfully generate the required examples using \texttt{kaist-ai/prometheus-8x7b-v2} \citep{KimEtAl:2024:Prometheus2}. 

\begin{figure}
    \centering
    \includegraphics[width=\linewidth]{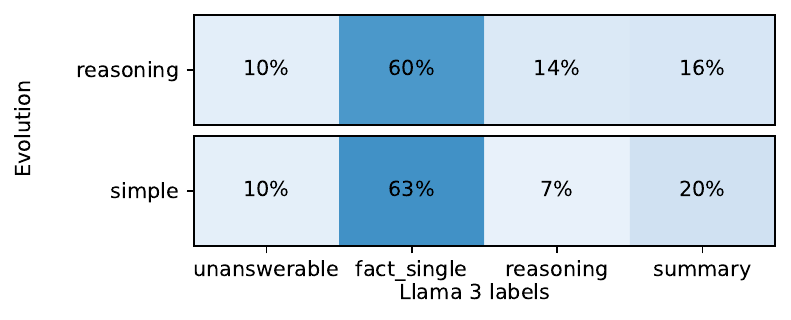}
    \caption{Distribution of \llama{3} labels for questions generated by Ragas using \emph{simple} and \emph{reasoning} evolutions}
    \label{fig:ragas_eval}
\end{figure}

\Cref{fig:ragas_eval} shows the result of passing the generated context-question pairs through our \llama{3}-based labeller described in \Cref{section:llm_labeler}. Note that \factsingle{} is over-represented in the output of both evolutions. In contrast, our models generate a more significant fraction of reasoning questions when asked to do so (see \Cref{fig:gen_labels}). Additionally, as expected, the \emph{simple} evolution produces a sizable portion of both \factsingle{} and \summary{} questions instead of being a \emph{pure} class with respect to our taxonomy. One can use our significantly cheaper fine-tuned model to generate a more balanced dataset than Ragas, as is evident from \Cref{fig:gen_labels}.

We also critiqued the Q\&A pairs generated by Ragas using our critiques and found the scores to be similar to our statement extraction method.

\end{document}